\documentclass[preprint]{aastex}
\newcommand{\be}{\begin{eqnarray}}
\newcommand{\ee}{\end{eqnarray}}
\newcommand{\beq}{\begin{equation}}
\newcommand{\eeq}{\end{equation}}
\def\simless{\mathbin{\lower 3pt\hbox
      {$\rlap{\raise 5pt\hbox{$\char'074$}}\mathchar"7218$}}} 
\def\simgreat{\mathbin{\lower 3pt\hbox
      {$\rlap{\raise 5pt\hbox{$\char'076$}}\mathchar"7218$}}} 

\newcommand{\vvec}{{\mbox{\boldmath $v$}}}
\newcommand{\kvec}{{\mbox{\boldmath $k$}}}
\newcommand{\avec}{{\mbox{\boldmath $a$}}}

\newcommand{\Bvec}{{\mbox{\boldmath $B$}}}
\newcommand{\Fvec}{{\mbox{\boldmath $F$}}}

\newcommand{\Evec}{{\mbox{\boldmath $E$}}}
\newcommand{\bvec}{{\mbox{\boldmath $b$}}}
\newcommand{\uvec}{{\mbox{\boldmath $u$}}}
\newcommand{\grad}{{\mbox{\boldmath $\nabla$}}}
\newcommand{\unit}{{\mbox{\boldmath $e$}}}
\newcommand{\bfdelta}{{\mbox{\boldmath $\delta$}}}

\newcommand{\calEvec}{{\mbox{\boldmath ${\cal E}$}}}

\newcommand{\Ptot}{P_*}

\begin{document}

\title{A Free, Fast, Simple and Efficient TVD MHD Code}
\author{ Ue-Li Pen$^{1}$, Phil Arras$^{2}$, ShingKwong Wong$^{1,3}$ }
\affil{$^1$ Canadian Institute for Theoretical Astrophysics, University
of Toronto \\
$^2$ Kavli Institute for Theoretical Physics\\
$^3$ Physics Department, National Taiwan University }
\authoraddr{$^1$ 60 St. George Street, Ontario M5S 3H8 Canada \\
arras@cita.utoronto.ca,pen@cita.utoronto.ca}


\begin{abstract}

We describe a numerical method to solve the magnetohydrodynamic (MHD)
equations. The fluid variables are updated along each direction using
the flux conservative, second order, total variation diminishing (TVD),
upwind scheme of Jin and Xin. The magnetic field is updated separately
in two-dimensional advection-constraint steps. The electromotive force
(EMF) is computed in the advection step using the TVD scheme, and this
same EMF is used immediately in the constraint step in order to preserve
$\grad \cdot {\bf B}=0$ without the need to store intermediate fluxes.
Operator splitting is used to extend the code to three dimensions, and
Runge-Kutta is used to get second order accuracy in time.  The advantages
of this code are high resolution per grid cell, second order accuracy
in space and time, enforcement of the $\grad \cdot \Bvec=0$ constraint
to machine precision, no memory overhead, speed, and simplicity. A 3-D
Fortran implementation less than 400 lines long is made freely available.
We also implemented a fully scalable message-passing parallel MPI version.
We present tests of the code on MHD waves and shocks.

\end{abstract}



\section{ Introduction }

Astrophysical fluids in which the magnetic field plays an important
role are common in nature. As a few examples, consider
magnetized interstellar gas, accretion disks, molecular
clouds, and jets.  With the advent of high speed computers and ever
improving MHD codes, considerable theoretical progress has been made
through numerical simulation of otherwise intractable problems.

A major challenge to solving flux conservative systems of equations,
such as ideal fluids and MHD, is the spontaneous development of shock
discontinuities.  Finite differencing across discontinuities leads to
divergences and instabilities.  Modern codes implement various aspects
of ``flux limiters'' \citep{Harten} to achieve stability near shocks
and second order accuracy away from shocks.  Recently, several shock
capturing methods which solve the MHD equations in flux conservative
form with upwind finite differencing have been developed. Enforcing
the $\grad \cdot \Bvec = 0$ constraint is key to the accuracy of these
codes Near shock fronts, derivatives are ill-defined, and the divergence
constraint can be maximally violated.  \citet{1988ApJ...332..659E}
first noted that this ``constrained transport" (CT) can easily be
enforced to machine precision by (1) defining the magnetic field at
cell faces instead of centers, and (2) using the same EMF, computed
on cell corners, to update the magnetic flux through each neighboring
face. Using CT and various shock capturing schemes, several groups (see
e.g., \citet{2000JCoPh.161..605T} for a review of different methods)
have now produced robust, efficient MHD codes.

The detailed algorithm used for the finite differencing varies
between the different groups. We do not give an exhaustive review of
the literature but rather compare only to widely used codes, or those
similar to ours.  The Zeus code \citep{1992ApJS...80..791S} partially
updates certain fluid and magnetic field quantities along Alfven, but not
magnetosonic, characteristics to avoid short lengthscale instabilities in
shear Alfven waves. \citet{1998ApJ...509..244R} use Harten's TVD method,
which evolves the fluid along all the characteristics by constructing
the linearized eigenvectors.  Common to these two methods is the need
to first compute the EMF over the whole grid, then perform a spatial
averaging of the EMF, and then update the magnetic field.

In this paper we implement the divergence constraint in a slightly different
way than previous investigators. We show that individual pieces of the EMF
can be used in advection-constraint steps, {\it without the need to store the
computed EMF's over the whole grid}. This gives us a sizeable savings in 
memory than if the EMF's were stored.  Furthermore, by using Jin and Xin's (1995)
``symmetric" \footnote{ This method is called symmetric since it
decomposes each fluid quantity into left and right moving parts, each
of which can be differenced in an upwind fashion.} method of computing
TVD fluxes, we can reduce the operations count relative to codes which
manifestly evolve the fluid along characteristics.

In section \ref{sec:equations} we review the MHD equations. In section
\ref{sec:method} we describe our numerical method. Tests of the code are
presented in section \ref{sec:tests}. Section \ref{sec:merits} contains
a discussion of the merits and drawbacks of the code. Section
\ref{sec:conclusion} contains the conclusions. We briefly review the
\citet{jinxin} method for solving one dimensional advection equations
in an appendix.

\section{ Equations }
\label {sec:equations}

The MHD equations expressing conservation of mass, momentum and  energy, as
well as
magnetic flux freezing are \citep{1984ecm.book.....L}
\beq
\partial_t \rho + \grad \left( \rho \vvec \right)  =  0
\label{eq:m}
\eeq
\beq
\partial_t \left( \rho \vvec \right) 
+ \grad \left( \rho \vvec \vvec + \Ptot \bfdelta
- \bvec \bvec \right)  = \rho \avec
\label{eq:p}
\eeq
\beq
\partial_t e + \grad \left[ (e+\Ptot) \vvec - \bvec \bvec \cdot \vvec \right]
 =  \rho \vvec \cdot \avec
\label{eq:e}
\eeq
\beq
\partial_t \bvec  =  \grad \times \left( \vvec \times \bvec \right)
\label{eq:b}
\eeq
\beq
\grad \cdot \bvec  =  0
\label{eq:divb}
\eeq
\beq
\Ptot  =  p + \frac{b^2}{2}
\label{eq:ptot}
\eeq
\beq
e  =  \frac{\rho v^2}{2} + \frac{p}{\gamma-1} + \frac{b^2}{2}.
\label{eq:ideal}
\eeq
Here $\rho$ and $e$ are the mass and (total) energy densities,
$\vvec$ is the velocity, $\Ptot$ is the total pressure, $p$ is the
gas pressure, $\bvec=\Bvec/\sqrt{4\pi}$ is the magnetic field in terms
of $\sqrt{4\pi}$, $\avec$ is an externally imposed acceleration, and
$\bfdelta$ is the Kronecker delta symbol.  In eq.\ref{eq:ideal} we have
used an ideal gas equation of state with internal energy $\varepsilon
=p/(\gamma-1)$,
where $\gamma$ is the ratio of specific heats. 
The infinite conductivity limit has
been used so that the electric field is $\Evec = - \vvec \times \Bvec
/c$. The electric force has been ignored since it is assumed that charge
separation is negligible on the scales of interest.

\section{ Numerical method}
\label{sec:method}

First we describe the update of the magnetic field in two dimensional
advection- constraint steps. We then briefly review the update of the
fluid variables along one dimension. Next we discuss how 
operator splitting and Runge-Kutta can be used to make the code second order accurate
in space and time. Finally, we discuss boundary conditions, fine tunings of the code,
and the parallel implementation.

\subsection{ solution of the induction equation in 2D} 

\begin{figure}
\plotone{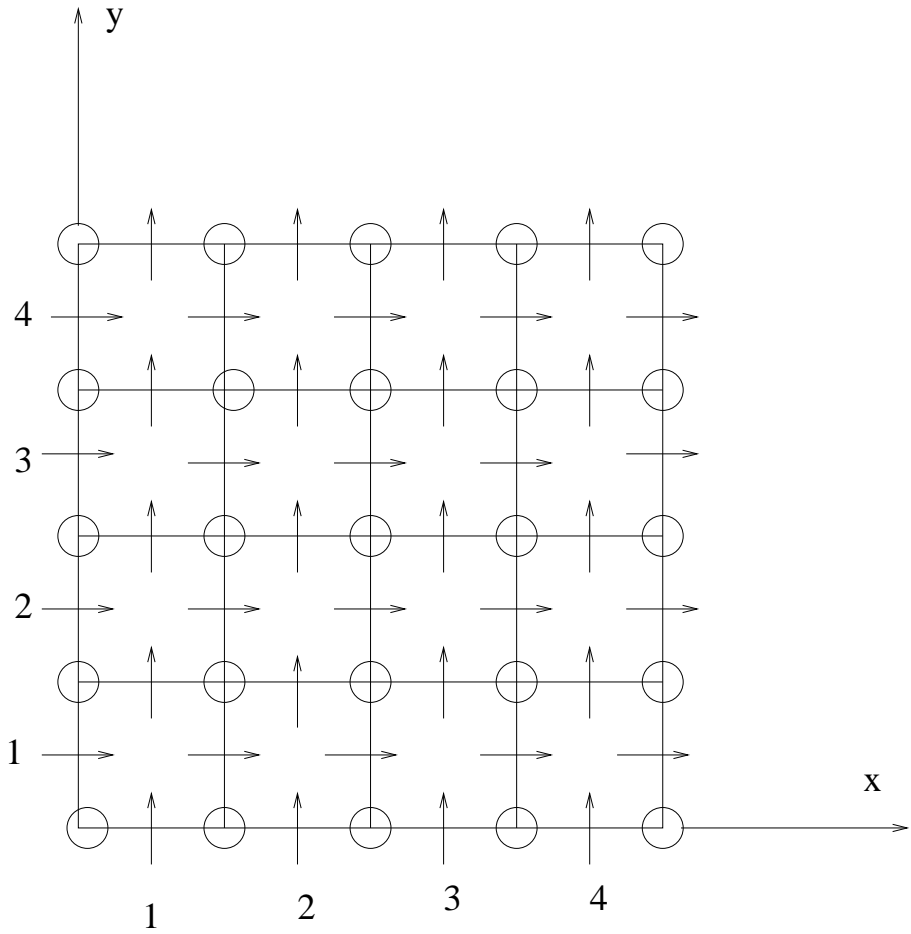}
\vspace{-2in}
\caption[]{ Position of variables on the grid. The vertical (horizontal)
arrows represent $b_y$($b_x$) respectively. The circles denote the position
of the electromotive force. }
\label{fig:grid}
\end{figure}

We use operator splitting to reduce the problem into a series of
smaller decoupled equations.  Alternating the order of operators in
the correct fashion allows one to achieve net second order accuracy.
In this prescriptions,  we hold the fluid variables fixed
to update the magnetic field.
The magnetic field is defined on cell faces (see fig.\ref{fig:grid})
in order to satisfy the $\grad \cdot \bvec=0$ to machine precision. Let
the cell centers be denoted by $(i,j,k) \equiv (x_i,y_j,z_k)$, and
faces by $(i\pm 1/2,j,k)$, $(i,j\pm 1/2,k)$, and $(i,j,k\pm 1/2)$,
etc. For convenience, let the cells have unit width.  The magnetic
field is then stored in arrays 
\be bx(i,j,k) & = & b^x_{i-1/2,j,k}
\nonumber \\ 
by(i,j,k) & = & b^y_{i,j-1/2,k} 
\nonumber \\ 
bz(i,j,k) & = & b^z_{i,j,k-1/2}.
\ee 
The flux out of cell $(i,j,k)$ is then 
\be [
\int d^3x \grad \cdot \bvec ]_{ijk} & = & b^x_{i+1/2,j,k}-b^x_{i-1/2,j,k}
\nonumber \\ & + &
 b^y_{i,j+1/2,k}-b^y_{i,j-1/2,k}
\nonumber \\ & + &
 b^z_{i,j,k+1/2}-b^z_{i,j,k-1/2}.
\ee 
Since the magnetic field is defined on cell faces, where the magnetic
flux in a cell is evaluated, it is possible to enforce $\grad \cdot
\bvec=0$ to machine precision \citep{1988ApJ...332..659E}.
 If we defined the magnetic field at some
other location, the divergence could only be kept to zero to truncation
error arising from interpolation to the faces.  The truncation errors
are necessarily large near shock discontinuites.
Next we describe a method
to evolve the field which preserves $\grad \cdot \bvec=0$, if it is
so initially.

When the induction equation is written out in spatial components, it is apparent that
the terms involving $\calEvec \equiv \vvec \times \bvec$
come in six pairs. For instance, the terms involving $v_y b_x$ are
\be
\partial_t b_x + \partial_y(v_y b_x) & = & 0
\nonumber \\
\partial_t b_y  & = &  \partial_x(v_y b_x).
\ee
The first equation is just the advection of $b_x$ along the $y$ direction, the
second equation is a {\it constraint} which enforces $\grad \cdot \bvec = 0$.
The key point to note here is that to enforce $\grad \cdot \bvec = 0$ we must use
the same EMF computed in the advection step during the constraint step; otherwise
$\grad \cdot \bvec = 0$ will only be zero up to truncation error.
We accomplish this by finding a second order accurate, upwind EMF $v_y
b_x$ for the advection step to update $b_x$, and then immediately use
this same EMF for the constraint step to update $b_y$. \citet{1998ApJ...509..244R}
first store the EMF's over the entire 3D grid, then average the EMF's,
and then update the field. We construct the EMF using Jin and Xin's (1995)
TVD method, which is described in the appendix. Note that the velocity
$v^y_{ijk}$ must be interpolated to the
same position as the magnetic field $b^x_{i-1/2,j,k}$ with second
order accuracy.

Jin and Xin's (1995) symmetric method introduces a ``flux freezing speed" $c$, which
must be greater than or equal to the maximum speed at which information
can travel. Since we are holding the fluid variables fixed, the flux
freezing speed for the advection-constraint equation is just $c=|v_y|$.

\subsection{ solution of the fluid equations in 1D }

Now we briefly describe the fluid update. A more complete discussion is given
in \citet{2003PASP..115..303T}. The magnetic field is held fixed, and
interpolated to grid centers with second order accuracy.
Let $\uvec=(u_1,u_2,u_3,u_4,u_5)=(\rho,\rho v_x,\rho v_y,\rho v_z,e)$
represent the volume averaged quantities positioned at the center of 
each cell. For advection along the $x$ direction, the Euler, continuity,
 and energy equations can be
written in flux conservative form as
\be
\partial_t \uvec + \nabla_x \Fvec & = & 0 
\label{eq:euler} 
\ee
where the flux vector is given by
\[ \Fvec = \left( \begin{array}{l} 
\rho v_x \\
\rho v_x^2 + \Ptot - b_x^2 \\
\rho v_x v_y - b_x b_y \\ \rho v_x v_z -b_x b_z \\
(e+\Ptot) v_x - b_x \bvec \cdot \vvec 
\end{array} \right) \]
and the pressure is determined by $p=(\gamma-1)(e-\rho v^2/2 - b^2/2)$.
We hold the magnetic field fixed during the fluid update, and interpolate $\bvec$
to cell centers for second order accuracy.

Eq.\ref{eq:euler} can be solved by symmetric TVD, described in the appendix.
The flux freezing speed is taken to be 
$c=|v_x|+(\gamma p/\rho + b^2/\rho)^{1/2}$, which is the maximum speed information
can travel \footnote{ We have taken the maximum speed of the fast magnetosonic wave
over all directions.}.

\subsection{ extension to 3D } 

Let the fluid update step for a time $\Delta t$ along $x$ be denoted
by ``fluidx", and update of $b_x$ along $y$ by ``bxalongy".  Operator
splitting requires us to apply each operator first in forward, and then in
reverse order to advance by two timesteps.  We implemented two versions.
In one, we advance forward using the sequence of operations: fluidx,
byalongx, bzalongx, fluidy, bxalongy, bzalongy, fluidz, bxalongz,
byalongz, and then the reverse byalongz, bxalongz, fluidz, bzalongy,
bxalongy, fluidy, bzalongx, byalongx, fluidx.  A second implementation,
used in the public version of this code, is to transpose the fluid
variables and spatial dimensions (see, e.g. \citet{nrf90} pg.984).
 This is only easily done if two of the
dimensions are equal.  Transposing has the benefit of high efficiency on
cache based computers, where we only need to read data in column order.
It is convenient to implement the 3D code by using a single routine
for the fluid update along the $x$ direction and a single routine for
advection of $b_y$ along $x$, and constraint of $b_x$ along y. The
order of the spatial indices is transposed to take account of the other
directions. The advantage of having one subroutine is that it can be
heavily optimized.

The time step is set by the fastest speed at which information travels over the
grid. Since the fluid update is more restrictive, the time step is set to be
\be
\Delta t & = & {\rm cfl}\ \left[ max(|v_x|,|v_y|,|v_z|)
+(\gamma p/\rho + b^2)^{1/2} \right]^{-1}
\ee
where cfl $\simless 1$ is generally set to cfl $\simeq 0.7$ for stability.

\subsection{ boundary conditions }

The standard boundary conditions such as periodic, continuous, or reflecting
can be easily enforced by specifying values of the variables in ``ghost zones"
adjacent to the physical grid.
These ghost zones are needed when interpolating $\vvec$ to cell faces,
$\bvec$ to cell centers, and in the one-dimensional advection routines. 

We have implemented the boundary conditions in two different ways. The first
method is to pad the grid with a large number ($\sim 6-10$) of extra cells at
each boundary. Both the on-grid and off-grid variables are evolved in time, but
so many extra cells are used that the boundary cells only need be updated once
per double time step. This method is useful for parallel implementations in which
buffer zones are used to represent a small number of cells in adjacent regions.
The second method to implement the boundary conditions requires that one write
specific routines for interpolation or derivatives which specify the off grid
values. This method requires less computation, and is preferable for serial
applications. We find it convenient to evolve ``extra"
values of the magnetic field variables. That is, we evolve $b^x_{i-1/2,j,k}$
for $j=1,...,n_y$ and $k=1,...,n_z$ {\it but} $i=1,...,n_x+1$. This is useful
for three reasons. First, this allows $\grad \cdot \bvec$ to be computed over all
cells. Second, to update $b^y_{n_x,j+1/2,k}$ the fluxes $v_y b_x$ are needed at
in the boundary cells (see fig.\ref{fig:grid}) with $i=n_x+1/2$.
Third, $\bvec$ can be interpolated to cell centers without the need to specify 
off grid values.

\subsection{ fine tunings of the code }
\label{sec:finetune}

Since the TVD limiters are nonlinear, sinusoidal waveforms can tend
to become ``clipped", or boxy-looking. We find that these nonlinear
distortions can be minimized by using constant flux freezing speed,
set to be the maximum along that advection line. Additional stability
can be gained by multiplying the flux freezing speed by a constant
multiplicative factor, although this increases the number of time steps
needed and makes the code more diffusive.  Empirically we find that
smoothing the velocity field which advects the magnetic field can lead
to less damping of the slow mode.

In production runs, we have found an occasional failure of the code when
the Courant condition is pushed too close to the limit.  In the operator
split approach, the time step is fixed at the beginning of a double
time step, and determined from the Courant condition at the beginning.
During the time step, this condition may change, leading to an instability
if it exceeds the initial constraint.  Our solution has been to be sufficiently
conservative using a choice of  cfl$\lesssim 0.7$.  A more efficient procedure
would be to measure the change in the Courant condition during the sweeps,
and use this as an indicator in subsequent time steps.  And should a
given sweep step be instable, one can always break it into two substeps.

\subsection{ Parallel implementation }

We have implemented a fully distributed version in MPI.  After a full
set of operators in one dimension, we update the buffer zones.  For
hydrodynamics, only 3 buffer cells are required.  The magnetic field
requires interpolation, and we use 16 buffer cells for magnetized
simulations.  A full three dimensional domain decomposition is
implemented, where we update the buffers in the appropriate direction
after each dimensional operator.  Since only large faces are
communicated, latency of communication is negligible, but signficant
bandwidth is required to move the buffer zones.  The communication is
performed asynchronously, and computations proceed during the
communication stage.  Within each node, OpenMP is used to utilize
multiple processors in a node without the overhead of buffer cells and
communications.

We tested the parallel implementation on the CITA McKenzie beowulf
cluster.  The main cluster has 256 nodes of dual Intel Pentium-4 Xeon
processors running at 2.4 Ghz, 1 GB of RAM, dual gigabit ethernet, and 80
GB of disk.  The networking consists of bristles of 16 machines with one
gigabit port connected to a switch.  The second gigabit port is used to
interconnect the bristles in a cubical layout.  The nominal bi-section
bandwidth is 128 Gbit/sec.  For cubical problems, the usable bandwidth is
higher when the domain decomposition is matched to the cube as it is for
our runs.  To minimize communications overhead, we
mapped the computational grid layout to coincide with the physical
network interconnect.  The largest problem that we have been able to
run in memory is $1400^3$ grid zones, which takes about 40 seconds per
double time step.  The fine grained OpenMP parallelism within each nodes
allows the code to benefit from the intra-node hyperthread speedup.
The code also fully vectorizes for the SSE2 parallel execution units.
Due to the large number of buffer cells required, about 1/3 of the
computation and memory are used by these buffers.  The operation count
of the van Leer limiter relaxing TVD algorithm is 33 floating point
operations per variable per time step.  The flux computation takes an
additional 7 operations averaged per variable, for a total count of 40.
Each double time step consists of 6 sweeps of 8 variables, or about 2000
floating point operations.  Our execution speed corresponds to a sustained
rate of over 200 Gflop on the cluster, which is about 5\% of
theoretical peak speed of 4.8 Tflop in single precision.

\section{ Tests }
\label{sec:tests}

In this section we present tests of the code on MHD waves and shocks.

For all the tests we use $\gamma=5/3$ and set the box size $L$ to be
equal in all directions. Period boundary conditions were used for the
wave tests, and continuous boundary conditions (all variables continuous
across the boundary) in the shock.  The wave tests are two-dimensional
while the shock tests are along one dimension.  The van Leer limiter
(see, e.g. \citet{2003PASP..115..303T}) and a constant freezing speed
and cfl=0.7 (see section \ref{sec:finetune}) were used throughout.

\subsection{ torsional Alfven Waves }

Torsional Alfven waves are exact nonlinear solutions of the
compressible MHD equations. In the absence of any perturbations or
noise, they should propagate without steepening, making them a good
test for numerical codes.
\footnote{ In the presence of any infinitesimal noise, Alfven waves
are unstable to decay into three other waves
\citep{1978ApJ...219..700G,1978ApJ...224.1013D}.  For large amplitude
waves the growth time becomes comparable to the Alfven wave
period. Hence care must be used in applying this test to very
nonlinear waves as noise arising from truncation error or the
nonlinear flux limiters may grow exponentially. }

We perform tests for four different resolutions $n_y=n_z=16,32,64,128$,
where $n_{y,z}$ are the number of grid points in the $y$ and $z$
directions.  Different fluid pressures are used corresponding to low and
high $\beta=c_s^2/b^2 =0.1,10$.  The exact solution we input to the code
is $\rho=1$, $e=p/(\gamma-1) + 0.5 + A^2$,
\be
b_x & = & A\cos(k(y+z) - \omega t)  \\
b_y & = & \frac{1}{\sqrt{2}} \left( 1+A \sin(k(y+z)-\omega t) \right) \\
b_z & = & \frac{1}{\sqrt{2}} \left( 1-A \sin(k(y+z)-\omega t) \right) \\
v_x & = & - A\cos(k(y+z) - \omega t)   \\
v_y & = & - \frac{A}{\sqrt{2}} \sin(k(y+z)-\omega t)  \\
v_z & = & \frac{A}{\sqrt{2}} \sin(k(y+z)-\omega t).
\label{eq:alfvensoln}
\ee
The wavenumber and frequency are $k=2\pi/L$ and $\omega=\sqrt{2}k$
for the lowest order mode.  We set the ratio of wave to background
field to be $A=0.1$ for the low gas pressure test, and $A=1$ for the
high gas pressure test.  As long as the thermal energy is larger than
the kinetic energy of the wave, we can also test very non-linear waves
\footnote{For sufficiently small $\beta \ll 1$ and large amplitude
$A \sim 1$, we found the code is susceptible to a short lengthscale
instability. The timescale over which this instability develops depends
strongly on $\beta$ and $A$, and also the number of grid points. It is
unclear to us whether or not this is the physical decay instability of
Alfven waves due to perturbations seeded by the truncation error.  }.

\begin{figure}
\plottwo{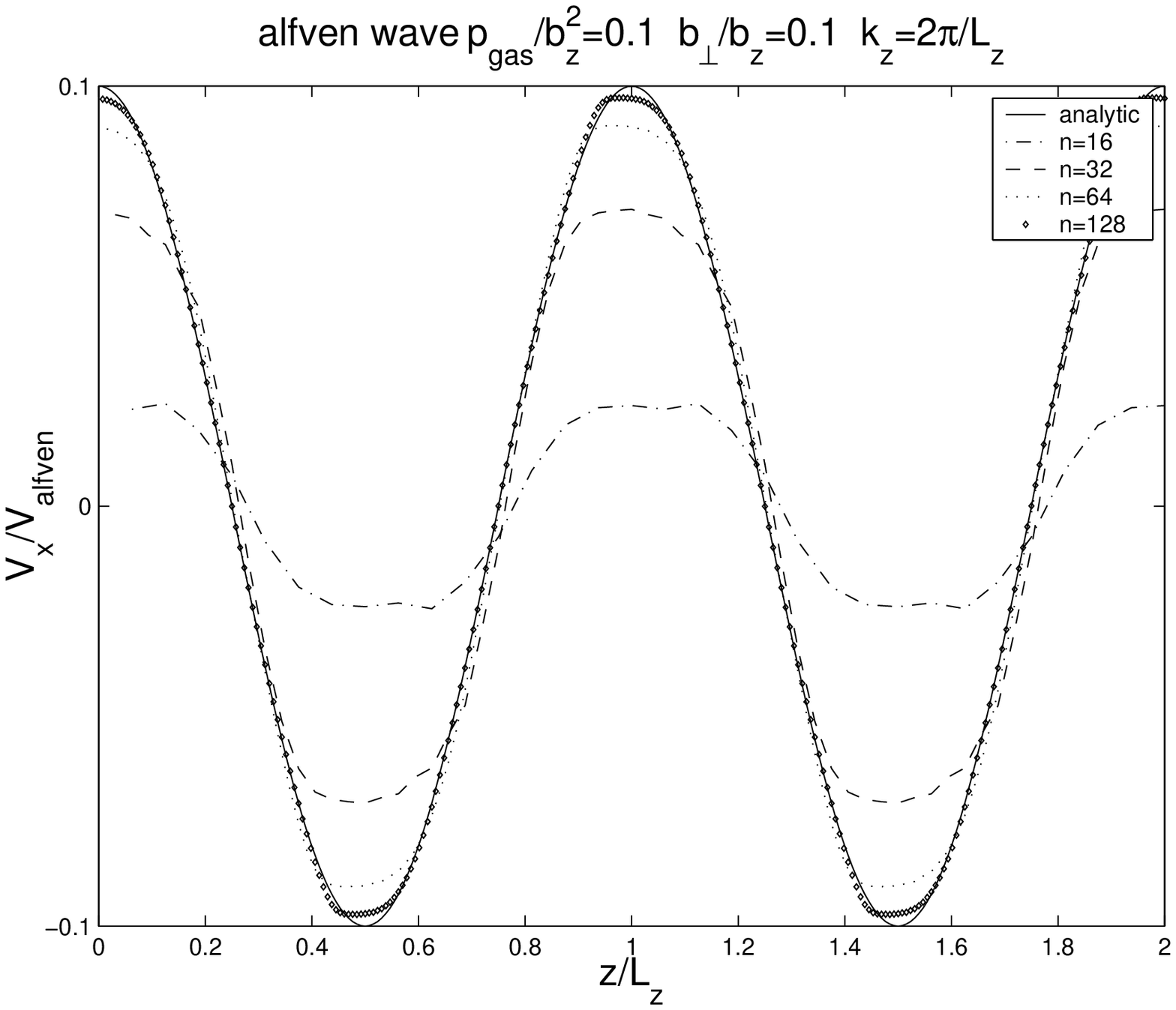}{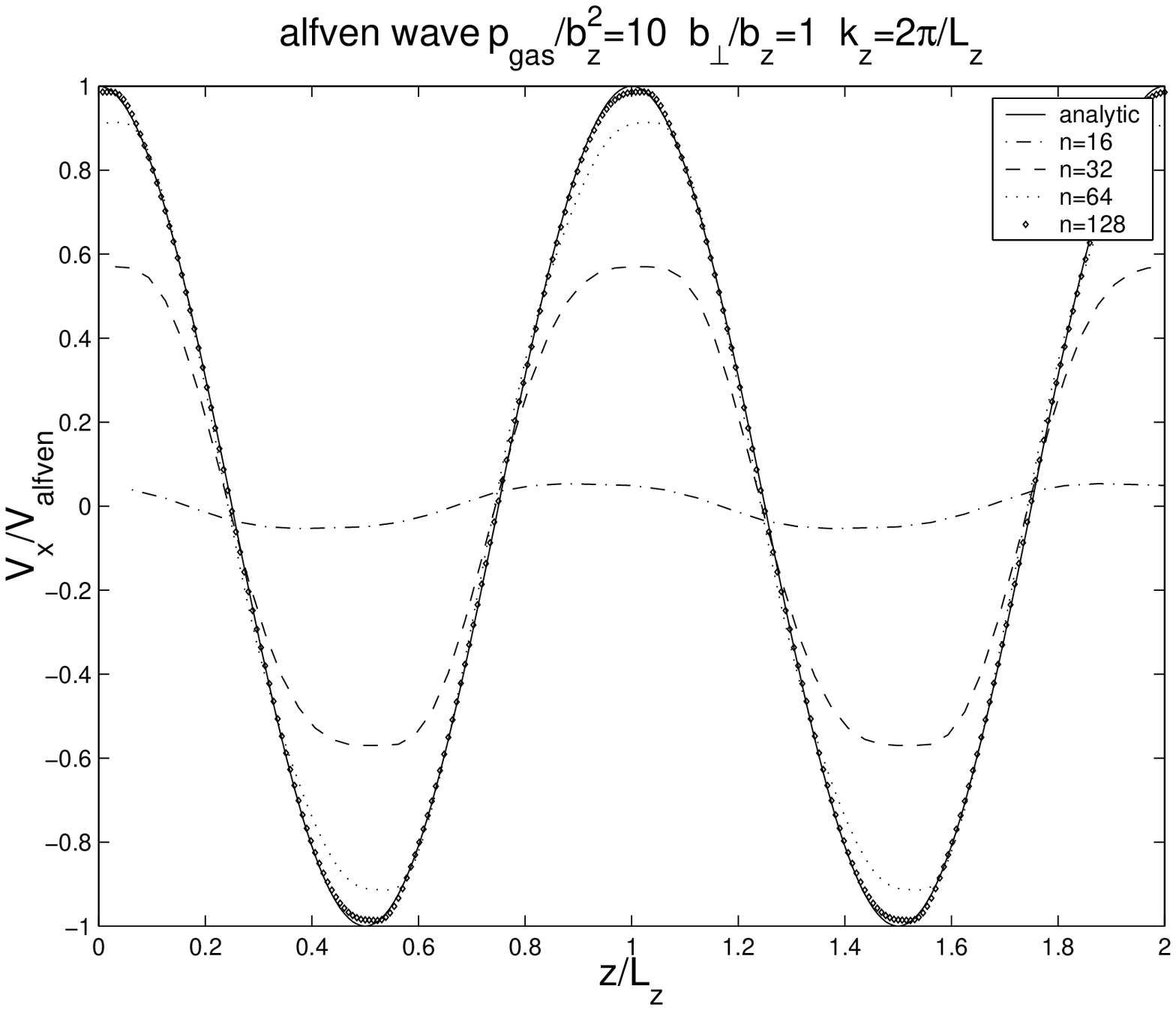}
\caption[]{ Torsional Alfven wave with amplitude $v_{wave}=0.1v_{Alfven}$
and $\beta=0.1$ (left) and $v_{wave}=1v_{Alfven}$
and $\beta=10$ (right) 
propagated for one wave period. The different
lines correspond to the analytic solution in eq.\ref{eq:alfvensoln} and
the code output for different numbers of grid points along the $z$
direction.  Each computational point is plotted twice over two
periods for clarity.}
\label{fig:alfven_lowbeta}
\end{figure}

The waves were propagated for one wave period. The result
is read along the $z$ axis and plotted against the exact solution in
fig.\ref{fig:alfven_lowbeta}. 
For clarity, we plotted two periods of
the wave (only one period was simulated).
The second order
convergence is apparent from the figures: as one doubles the number of
grid cells, the errors goes down by a factor of 4.
Even the significantly non-linear solutions
are well behaved and also converge quadratically.

\subsection{magnetosonic waves}

\begin{figure}
\plottwo{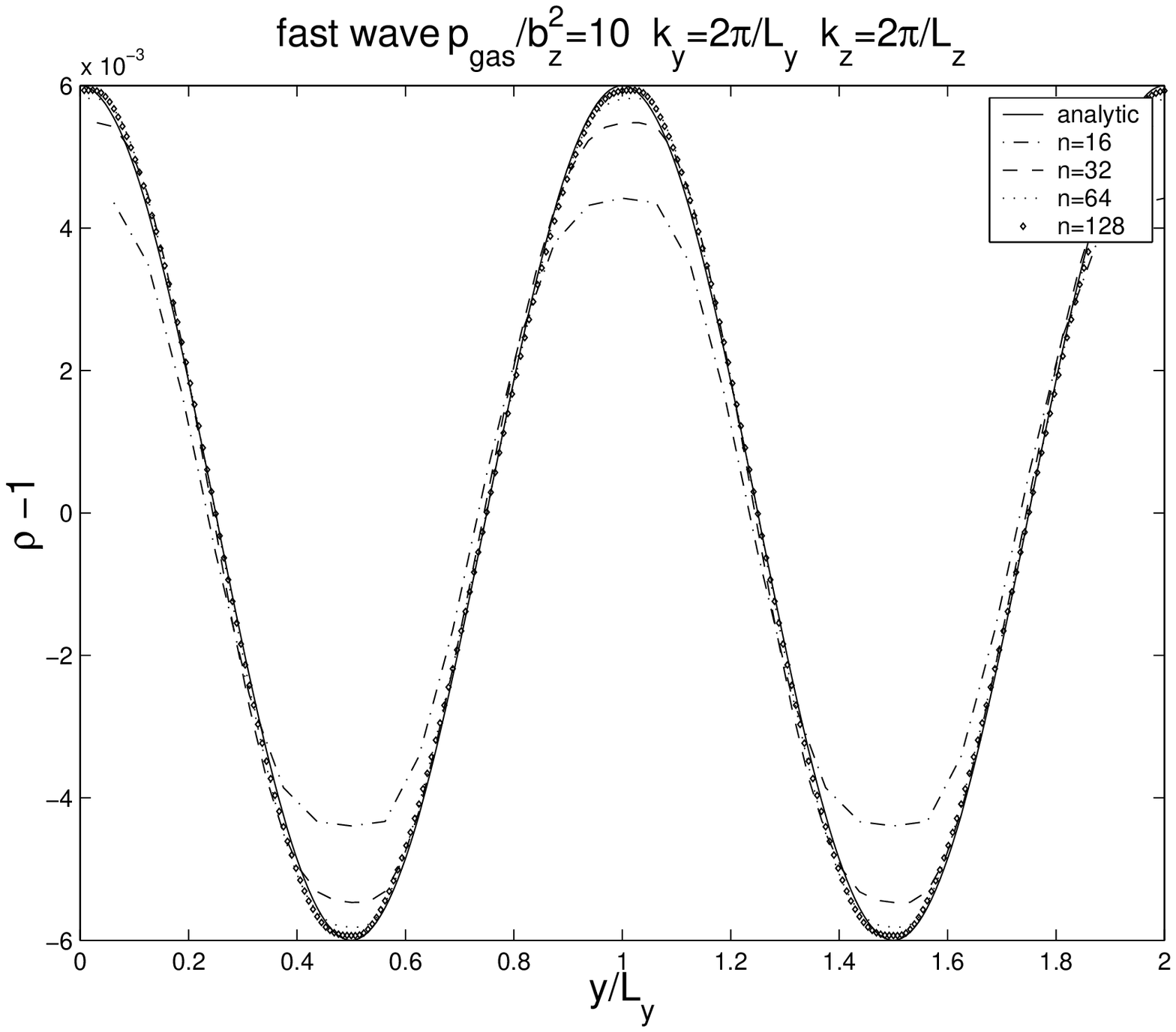}{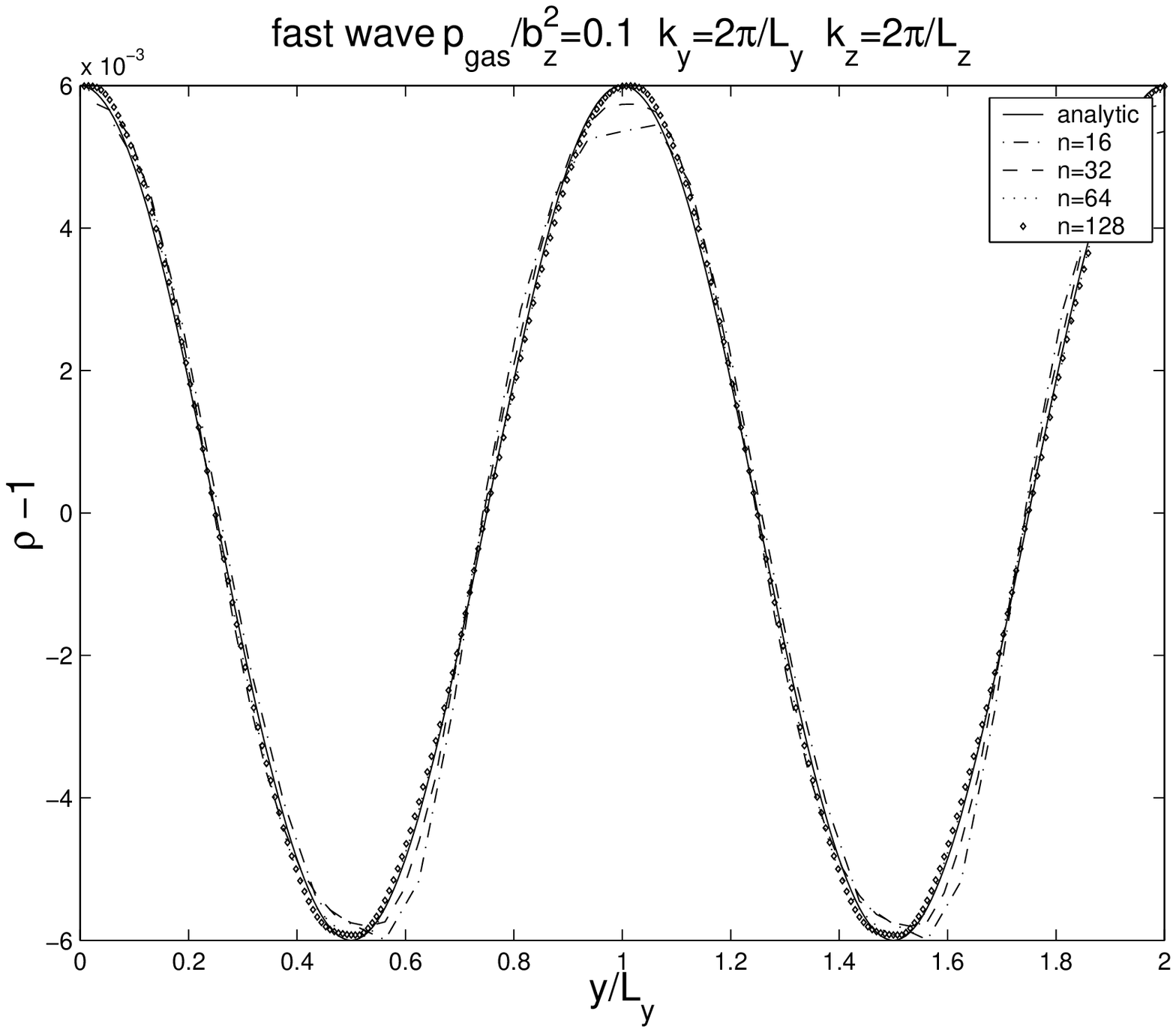}
\caption{ A fast magnetosonic wave with $k_yL=k_zL=2\pi$
and $p_{gas}/b_z^2=10,0.1$ on the left and right
panels, respectively, evolved for one wave period.  The field values
along the y axis are plotted for two periods, so each computational
cell is plotted twice.}
\label{fig:fast}
\end{figure}

\begin{figure}
\plottwo{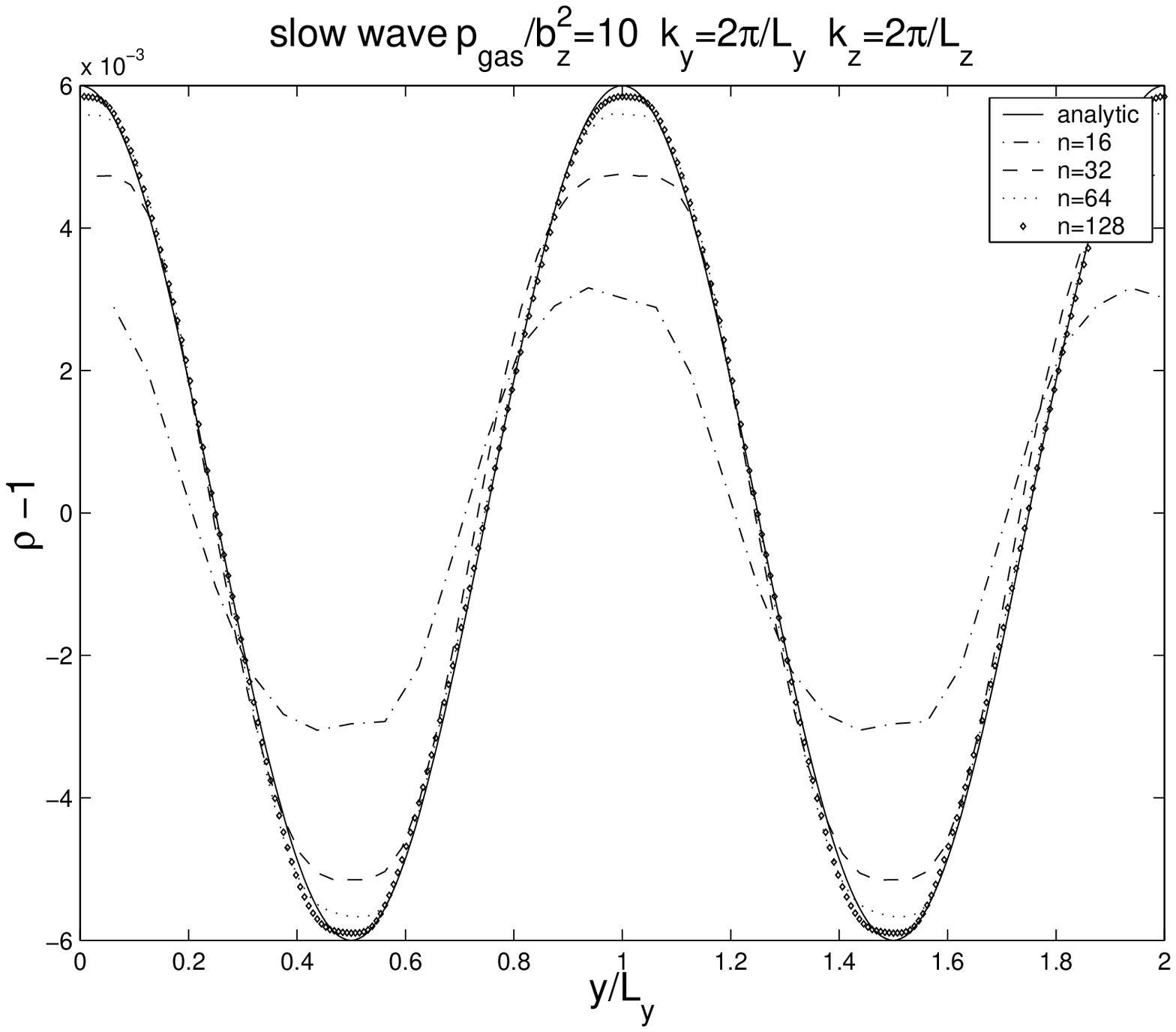}{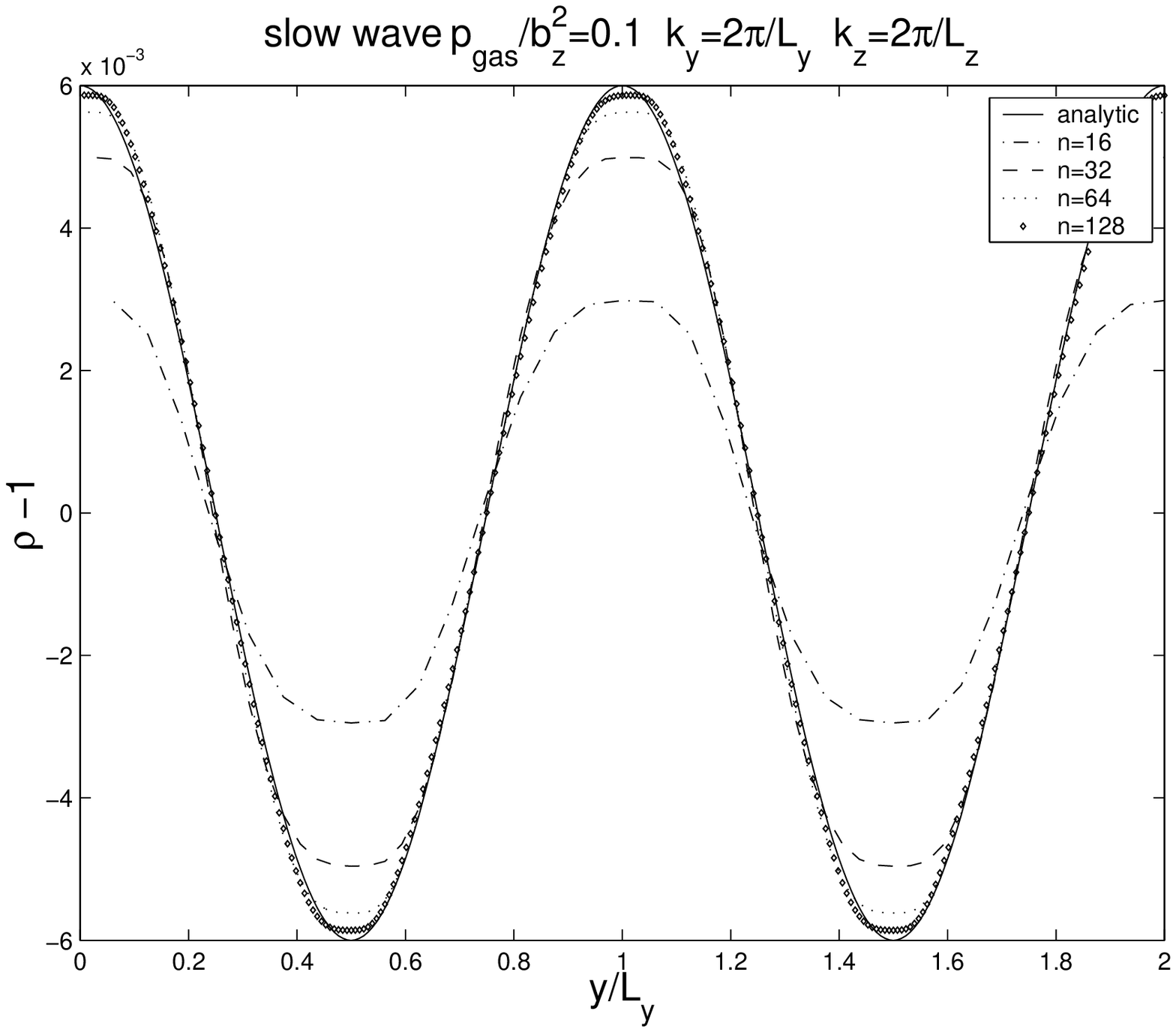}
\caption{ A slow magnetosonic wave with $k_yL=k_zL=2\pi$
and $p_{gas}/b_z^2=10.0,0.1$ on the left and right
panels, respectively, evolved for one wave period.  }
\label{fig:slow}
\end{figure}

We have tested magnetosonic waves in the linear regime $\delta p
\ll p$ for $\beta=0.1$ and $10$.  The background magnetic field is
$\bvec=\unit_z$, background pressure is $p=\beta$, and the wavevector
is ${\bf k}=k({\bf e}_y + {\bf e}_z)$, where $k=2\pi/L$ and $L=L_y=L_z$.
The exact solution we plug into the code is
\be
\delta p & = & p A \cos(k(y+z)-\omega t)
\nonumber \\
\rho & = & 1 + \frac{\delta p}{c_s^2}
\nonumber \\
\delta \vvec & = & \frac{\omega}{\omega^2-\omega_b^2} \left( 
\kvec - \frac{\omega_b^2}{\omega^2} \hat{\bvec} \hat{\bvec} \cdot \kvec\right) \frac{\delta p}{\rho}
\nonumber \\
\delta \bvec & = & \frac{k^2}{\omega^2-\omega_b^2} \left(
\bvec - \hat{\kvec} \hat{\kvec} \cdot \bvec \right) \frac{\delta p}{\rho}
\ee
where $c_s^2=\gamma p/\rho$, $\omega_b^2=k^2b^2/\rho$, and $A \ll
1$ is the wave amplitude.  The fast and slow mode frequencies are
given by $\omega^2_{f,s} =.5(\omega_b^2+\omega_s^2) \pm 0.5 \left(
(\omega_b^2+\omega_s^2)^2 - 4\omega_s^2 \omega_a^2 \right)^{1/2}$ where
$\omega_s^2=c_s^2k^2$ and $\omega_a^2=(\kvec \cdot \bvec)^2/\rho$. We
evolved the waves for one period. The
results for the fast and slow waves are shown in figures \ref{fig:fast}
and \ref{fig:slow}, respectively. 
The slow wave is subject to substantial
diffusion as compared to the fast wave since its frequency is so much
lower for these extreme values of $\beta$.

\subsection{ the MHD Riemann problem }

These 1d tests of the code involve a shock tube along the
x-axis, as in fig.2a,b of \citet{1998ApJ...509..244R}. We used
continuous boundary conditions and 1024 grid points for both tests.

\begin{figure} 
\plotone{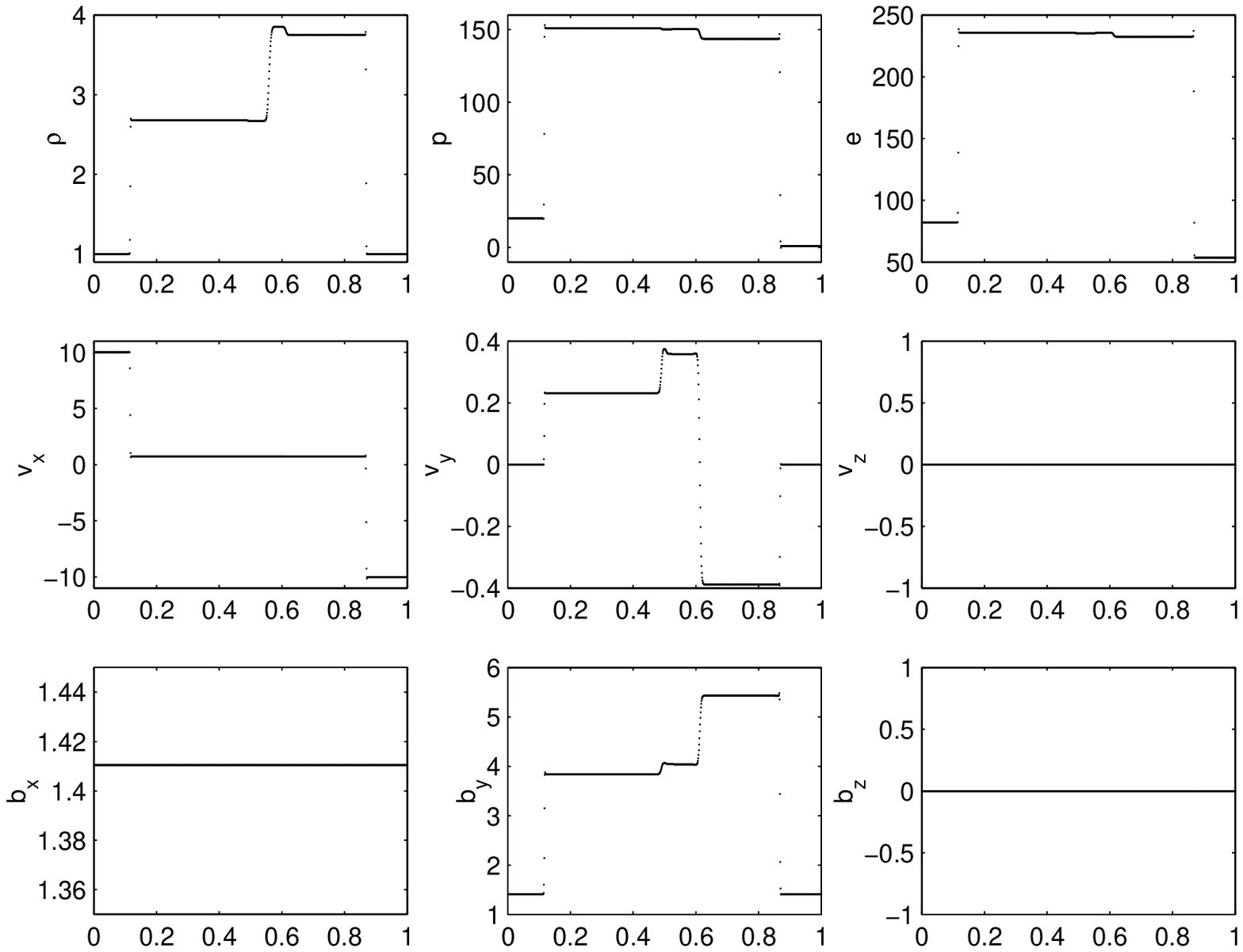} 
\caption[]{ Shock tube test along the x-axis with velocities and magnetic
field along the x and y directions.  }
\label{fig:shock1} 
\end{figure}

The initial conditions for fig.\ref{fig:shock1}
are $(\rho,v_x,v_y,v_z,p,b_x,b_y,b_z)=
(1,10,0,0,20,5/\sqrt{4\pi},5/\sqrt{4\pi},0)$ for the left side and
$(1,-10,0,0,1,5/\sqrt{4\pi},5/\sqrt{4\pi},0)$ for the right side.
The code is run for a time $0.08 L$. The result agrees well with fig.2a
of \citet{1998ApJ...509..244R}. The following features can be seen. The
steep discontinuities at $x\sim 0.1$ and $x\sim 0.85$ are fast shock
fronts where the incoming flow converts its kinetic energy into thermal
energy and compresses the transverse field $b_y$. As new matter falls
on, this shock is regenerated and maintains it's steep profile as it
moves outward. At $x\simeq 0.6$ and $x \simeq 0.5$ are a slow shock and
slow rarefaction, respectively. The slow shock again compresses the fluid
but decreases the transverse field. At $x\simeq 0.55$ the two phases of
the initial gas configuration with different entropies form a contact
discontinuity. Pressure, magnetic field and velocity are continuous
while density and thermal energy experience a discontinuity. This
discontuity moves rightward across the grid, and the TVD advection of such
discontinuities results in some smearing or diffusion of the structure.
No physical mechanism steepens this contact discontinuity once it smears,
and a slow numerical diffusion is visible in this, and all generic TVD
codes which do not introduce explicit contact steepeners.  There are
no significant oscillations. Both our and \citet{1998ApJ...509..244R}'s
solution have a slight overshoot in some variables in the first postshock
cell, but this effect does not persist onto subsequent cells.

\begin{figure}
\plotone{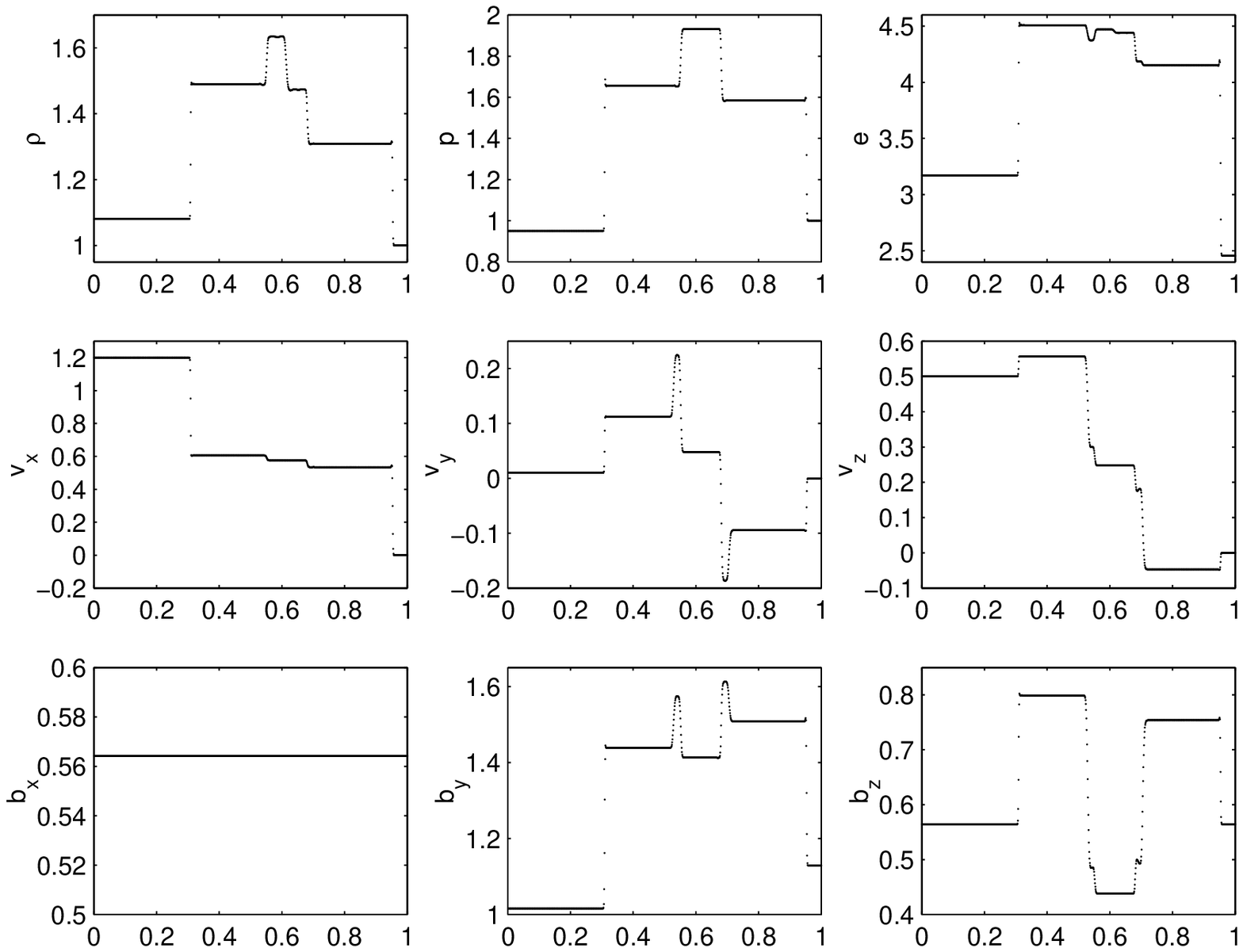}
\caption[]{ Shock tube test along the x-axis with velocities and magnetic  
field in all three directions. }
\label{fig:shock2}
\end{figure}

The initial condition for fig.\ref{fig:shock2} has velocity and magnetic field
components in all directions, and hence exhibits additional structures
such as rotational discontinuities. The values
are $(\rho,v_x,v_y,v_z,p,b_x,b_y,b_z)$ = $(1.08,1.2,0.01,0.5,
0.95,2/\sqrt{4\pi},3.6/\sqrt{4\pi},2/\sqrt{4\pi})$ on the left side and
$(1,0,0,0,1,2/\sqrt{4\pi},4/\sqrt{4\pi},2/\sqrt{4\pi})$ on the right
hand side. The code is run for a time $0.2 L$.  The results again agree
with \citet{1998ApJ...509..244R}. The following features may be seen:
fast shocks at $x\simeq 0.3$ and $0.9$, rotational discontinuity at 
$x\simeq 0.53$ right next to a slow shock at $x \simeq 0.55$, 
contact discontinuity at $x \simeq 0.6$,
slow shock and rotational discontinuity at $0.68$ and $0.70$ respectively.

The shock tests have two basic types of structures: self-steepening shock
fronts and non-evolutionary contact and rotational discontinuities.
For the active shock fronts, fig.\ref{fig:shock1} and \ref{fig:shock2}
show the shocks extend over leading to a nominal shock resolution of around 2 grid cells.
The internal contact discontinuities arise from the discontinuities in
the initial conditions, and diffuse numerically as they advect over the
grid.  At discontinuities, the solution is non differentiable which is in
general a challenge to numerical schemes.  In the TVD approach, the scheme
drops to first order accuracy, with some associated diffusivity.  This is
discussed in more detail in \citet{2003PASP..115..303T}.  The shock fronts
are self steepening, so the first order diffusivity is less noticable.

\section{ Discussion of the Merits and Drawbacks of Our Numerical Scheme }
\label{sec:merits}

The code described in this paper differs from its predecessors ( TVD
method for shock capturing, enforcement of $\grad \cdot \bvec=0$ to
machine precision ) in two main respects. First, we solve the induction
equation in 2D advection-constraint steps without storing intermediate
fluxes over the entire grid. Storage of the fluxes would require a $3n^3$
array, nearly half the memory used for the basic variables.  Second, we
use Jin and Xin's (1995) method to implement TVD, which requires only $\sim
30$ floating point operations per grid cell per time step per variable.
The benefits of these two methods are high resolution per grid cell,
low operations count, and simplicity of coding (the public version of
the 3D code is only 400 lines long.)

The price of separating the fluid and magnetic field updates into two
steps, rather than updating all variables at once, is that the coupling
between the velocity and magnetic field may be relatively weak compared to
other methods. It was this point which led \cite{1992ApJS...80..791S}
to use an update along Alfven characteristics. However, our tests on both
linear and nonlinear waves, as well as shocks, seem to indicate that in most
circumstances the code performs rather well and no instabilities arise.

There are regimes in which we know the current code to be inaccurate or
unstable. The generic setting is one where the characteristic families
have very different velocities. These can occur at low $\beta$ for large
amplitude Alfven waves, or highly supersonic flow with weak embedded
shocks. For such regimes, customized modifications to the algorithm
may be necessary (for the high mach number regime see, e.g. Trac and 
Pen 2003, in preparation; for the low $\beta$ case see Turner, Stone,
Krolik, and Sano, in preparation). Also, since Jin and Xin's (1995) method
does not explicitly evolve the fluid variables along characteristics,
our code may be more diffusive for low frequency waves when the ratio
of fast and slow wave speeds is large (either large or small $\beta$.)

\section{Conclusions}
\label{sec:conclusion}

We have presented the algorithm and tests for a simple and robust
MHD code which incorporates all features of modern high resolution
shock capturing. It is second order accurate away from extrema,
requires no memory overhead beyond storing the fluid variables,
optimizes easily to many computer architectures, and offers simplicity
in the coding. We have tested this code on linear and nonlinear MHD waves
as well as shocks. The single processor code can be freely downloaded at
{\begin{verbatim} http://www.cita.utoronto.ca/~pen/MHD \end{verbatim} }. 
We have also implemented a parallel version, which scales well on very
large commodity beowulf clusters.

\acknowledgements
We thank Neal Turner for several useful discussions. Phil
Arras is an NSF Astronomy and Astrophysics Postdoctoral Fellow. 
Computing resources were provided by the Canada Foundation for
Innovation.  SW is supported by the Taiwan NSC.

\appendix

\section{ advection in one dimension }

Here we review Jin and Xin's (1995) solution of the advection equation. We
will focus on a scalar equation, but extension to a vector equation is
straightforward. A more detailed discussion as well as code has recently been
published by \citet{2003PASP..115..303T}. 

The advection equation for a quantity $u$ with flux $f$ is 
\be
\partial_t u + \partial_x f = 0.
\ee
Jin and Xin's (1995) symmetric method is to define the new variable $w=f/c$
and equations for $u$ and $w$
\be
\partial_t u + \partial_x(cw) & = & 0
\nonumber \\ 
\partial_t w + \partial_x(cu) &= & 0.
\ee
These equations can be written in terms of left and right moving variables
by defining $u_r=(u+w)/2$ and $u_l=(u-w)/2$. These variables satisfy the 
equations
\be
\partial_t u_r + \partial_x(c u_r) & = & 0
\nonumber \\
\partial_t u_l - \partial_x(c u_l) & = & 0,
\label{eq:lr}
\ee
which describe information propagating to the right and left,
respectively. 

To solve eq.\ref{eq:lr} over a full time step with second order accuracy,
we first advance $u_r$ and $u_l$ over a half time step using the first
order upwind donor cell formula. These values are then used to construct
a second order accurate upwind flux using any of the known nonlinear
TVD limiters such as minmod, Van Leer, or superbee. Finally, given 
the updated values for $u_r$ and $u_l$, we reconstruct $u=u_r+u_l$.

For stability, the value of the flux freezing speed $c$ must be chosen 
larger than the speed at which information propagates. As discussed in the
text, we set $c=|v|$ when advecting the magnetic field, and $c={\rm cfl}(
|v|+(\gamma p/\rho + b^2/\rho)^{1/2} )^{-1}$.

How can one relate TVD to the ``artificial viscosity" schemes? These
schemes add in a nonlinear viscosity term in order to prevent
instabilities, as well as damp away oscillations which may occur near
disconintuities. However, this viscosity tends to prevent the formation
of discontinuities on scales of order one cell, severely degrading
the resolution of the simulation. TVD may be viewed as a strongly
nonlinear flux limiter which adds just enough diffusion to prevent
numerical instabilities.  TVD can often capture shocks in only one or
two cells. Away from discontinuities, maxima or minima, TVD is second
order in space, but at a maxima it is only first order.

\bibliography{apj-jour,ref}

\begin{thebibliography}{11}
\expandafter\ifx\csname natexlab\endcsname\relax\def\natexlab#1{#1}\fi

\bibitem[{{Derby}(1978)}]{1978ApJ...224.1013D}
{Derby}, N.~F. 1978, \apj, 224, 1013

\bibitem[{{Evans} \& {Hawley}(1988)}]{1988ApJ...332..659E}
{Evans}, C.~R. \& {Hawley}, J.~F. 1988, \apj, 332, 659

\bibitem[{{Goldstein}(1978)}]{1978ApJ...219..700G}
{Goldstein}, M.~L. 1978, \apj, 219, 700

\bibitem[{{Harten}(1983)}]{Harten}
{Harten}, A. 1983, J. Comp. Phys, 49, 357

\bibitem[{{Jin} \& {Xin}(1995)}]{jinxin}
{Jin}, S. \& {Xin}, Z. 1995, Comm. Pure and Applied Math., 48, 235

\bibitem[{{Landau} \& {Lifshitz}(1984)}]{1984ecm.book.....L}
{Landau}, L.~D. \& {Lifshitz}, E.~M. 1984, {Electrodynamics of continuous
  media} (Pergamon Press, Sao Paulo, 1984)

\bibitem[{{Press} {et~al.}(1996){Press}, {Teukolsky}, {Vetterling}, \&
  {Flannery}}]{nrf90}
{Press}, W.~H., {Teukolsky}, S.~A., {Vetterling}, W.~T., \& {Flannery}, B.~P.
  1996, {Numerical recipes in Fortran 90. The art of Parallel Scientific
  Computing} (Cambridge: University Press, |c1996, 2nd ed.)

\bibitem[{{Ryu} {et~al.}(1998){Ryu}, {Miniati}, {Jones}, \&
  {Frank}}]{1998ApJ...509..244R}
{Ryu}, D., {Miniati}, F., {Jones}, T.~W., \& {Frank}, A. 1998, \apj, 509, 244

\bibitem[{{Stone} \& {Norman}(1992)}]{1992ApJS...80..791S}
{Stone}, J.~M. \& {Norman}, M.~L. 1992, \apjs, 80, 791

\bibitem[{{T{\' o}th}(2000)}]{2000JCoPh.161..605T}
{T{\' o}th}, G. 2000, Journal of Computational Physics, 161, 605

\bibitem[{{Trac} \& {Pen}(2003)}]{2003PASP..115..303T}
{Trac}, H. \& {Pen}, U. 2003, \pasp, 115, 303

\end{thebibliography}
\bibliographystyle{apj}

\end{document}